\documentclass[]{spie}  

\usepackage[]{graphicx}
\usepackage[]{amsmath}
\usepackage{rotating}
\usepackage{amssymb}

\def\eps@scaling{1.0}%
\newcommand\epsscale[1]{\gdef\eps@scaling{#1}}%

\newcommand\plotone[1]{%
 \centering 
 \leavevmode 
 \includegraphics[width={\eps@scaling\columnwidth}]{#1}%
}%
\newcommand\plottwo[2]{%
 \centering 
 \leavevmode 
 \columnwidth=.45\columnwidth 
 \includegraphics[width={\eps@scaling\columnwidth}]{#1}%
 \hfil 
 \includegraphics[width={\eps@scaling\columnwidth}]{#2}%
}%

\title{Results from the GMT Ground-Layer Experiment at the Magellan Telescopes}

\author{A. Athey\supit{a}, S. Shectman\supit{a}, M. Phillips\supit{b}, J. Thomas-Osip\supit{b}
\skiplinehalf
\supit{a}Carnegie Observatories, Pasadena, CA 91101-1209, USA; \\
\supit{b}Las Campanas Observatory, Carnegie Observatories, Casilla 601, La Serena, Chile; \\
}

\begin{document} 
  \maketitle

\begin{abstract}
We present results from our two year study of ground-layer turbulence
as seen through the 6.5-meter Magellan Telescopes at Las Campanas
Observatory. The experiment consists of multiple, moderate resolution,
Shack-Hartmann wavefront sensors deployed over a large 16 arcminute
field. Over the two years of the experiment, the ground-layer turbulence
has been sampled on eleven nights in a variety of seeing and wind
conditions. On most nights the ground-layer turbulence contributes 10\% 
to the total visible-band seeing, although a few nights
exhibit ground-layer contributions up to 30\%. We present the ground-layer
turbulence on the sampled nights as well as a demonstration of
its strength as a function of field size. This information is combined
with data from a MASS-DIMM seeing monitor adjacent to the Magellan
Telescopes to infer the annual ground-layer contribution to seeing
at Las Campanas. 
\end{abstract}


\section{Introduction}
\label{sect:intro}
The concept of ground-layer adaptive optics correction (GLAO) is to make
a modest improvement in image quality over a large field of view (many 
arcmin) by compensating for the turbulence in the lowest layers
of the atmosphere.  GLAO is complementary to full-atmosphere adaptive 
optics correction (AO), which delivers diffraction-limited imaging, 
but only over small fields of view ($<$ 1 arcmin). 

Recent studies of the atmosphere at astronomical sites have been conducted
using a variety seeing monitors, and also with balloons carrying micro-thermal 
sensors \cite{1994A&A...284..311V,1998A&AS..130..141K,2001A&A...369..364A,MASS_3,2003RMxAC..19...23S,2003SPIE.4839..466W}.
A number these studies provide measuremants of both the total seeing and the
distribution of turbulence as a function of altitiude.
The new results suggest that typically more than half of the turbulent power 
in the atmosphere occurs at low altitude ($\lesssim$ 1.0 km). 
This ground-layer is thought to arise from the interaction of moving air with local
topography, which may differ from site to site depending
on wind direction and ground contour upstream of the telescope.

The presence of a strong, low altitude turbulence layer provides an opportunity
to improve the seeing using an AO system with modest operating parameters 
\cite{2000SPIE.4007.1022R}.  Near the ground, the wind speed
is low so the crossing time for a turbulent cell is typically on the
order of ten milliseconds or more.  Because the effective height of the ground layer is 
largest for the largest turbulent scales, a GLAO system favors correction of the 
lowest order modes.  For atmospheric layers near the telescope, the isoplanatic angle
will be large. 
Thus, a GLAO system might operate at low frequency (100
Hz) and low order ($<$50 modes on a 6.5-m telescope) while significantly
improving the image quality over a large field of view (many arcmin).  
Initial GLAO models predicted a gain of a factor of two in image size over 
fields of 10 arcmin or more, resulting in a factor of 
four gain in sensitivity for background limited observations 
\cite{2000SPIE.4007.1022R,2003SPIE.4839..673T}. The potential
of GLAO correction is evident, however most of the work to date has
consisted of modeling based on a few atmospheric measurements.  Only limited
amounts of data have been taken at the telescope.

Although several on-going studies reveal at least some information about
ground-layer turbulence \cite{2004SPIE.5490..758W,2005ApJ...634..679L},
none of the current experiments provides adequate empirical measurements of the 
parameters and performance of a GLAO system.  Early laser-tomography experiments, 
for example, are restricted to smaller fields of view.  Since in any event different 
observatory sites might be expected to exhibit different ground-layer conditions, 
we have chosen to conduct an experiment to directly measure
the effect of ground-layer turbulence at the Magellan Telescopes.  This experiment 
was conducted as part the Giant Magellan Telescope (GMT) project, in order to study 
the implications of GLAO for the design of GMT and its instrumentation.

\section{GLAO Instrument/Experiment Description} 

We have developed a system of multiple Shack-Hartmann wavefront sensors
to investigate the ground-layer of the atmosphere as seen through
one of the 6.5 meter Magellan telescopes. Up to four separate wavefront sensors are 
attached to a machined plate in the telescope focal plane at the locations of stars 
in a few bright asterisms (both real clusters and random groupings) selected at 
convenient locations around the sky. 
The cameras operate at a frame rate of 100 Hz.  The separations between stars range 
from a minimum of 4.5 arcminutes, set by mechanical interference between adjacent units, 
to 16 arcminutes, set by the field of view of the standard Magellan guider package and
the size of the machined plate. The system uses off-the-shelf components to minimize cost
and development time at the expense of sensitivity. 

The optics consist of a stock achromatic field lens and a lenslet array which produces 
multiple images of the star directly on the detector. The A.O.A. Inc. lenslet array has 
a pitch of 203$\mu$m and a focal length of 5.8 mm.  The optics sample the pupil in an 
11x11 square grid, resulting 90 usable subapertures which are inside the perimeter of 
the primary mirror and unobscured by the shadow of the secondary.  The subapertures are 
59 cm square at the primary mirror, which is somewhat larger than the coherence scale so
there is some seeing within a subaperture.  The original choice of subaperture size was 
was motivated by the idea that it might be possible to implement a functional low-order, 
wide-field GLAO system based on natural guide stars.

The simplest algorithm for isolating the wavefront error contributed by the ground layer 
is just to average the outputs of multiple wavefront sensors distributed around 
the field of view.  The analysis is simplified if the individual subapertures are accurately 
registered to identical pupil positions in all of the wavefront sensors.  This requires 
both rotational alignment of the lenslet arrays and cameras, as well as x-y alignment of 
the lenslet arrays on the pupil.  In order to accomplish the x-y alignment, the lenslet array 
is deliberately offset from the exact focus of the field lens.  Each wavefront sensor is 
mounted on an x-y stage.  Adjustment of the x-y position of the stage changes the positions of 
star images on the camera, but because the lenslet array is not quite at the focus of the 
field lens, the adjustment also changes the registration of the lenslet array on the pupil.
In our wavefront sensors, a 1.0 arcsecond motion of the star corresponds to a 75$\mu$m motion on 
the detector.  The spacing between subapertures is 2.6 arcseconds.  However the motion required 
to change the alignment of the subapertures on the pupil is much larger: a displacement of 5 
arcseconds will move the subaperture position about one-half of the distance between
subapertures.  Strictly speaking, the subaperture alignment should be referenced to the conjugate 
height of ground layer.  This was accomplished by mounting a special pupil mask near the 
center of the Magellan secondary mirror.  Since the telescope is Gregorian the conjugate image
of the secondary is located about 60 m above the primary.  The wavefront sensor mounts include 
an additional motion along the z-axis which is used to adjust the focus of the wavefront sensor 
(or equivalently the pitch of the Shack-Hartmann array on the detector) in order to compensate 
for the curvature of the telescope focal plane.

The cameras are commercial units purchased from Basler Vision Technologies.
The Basler A602f cameras use a Micron CMOS detector (MT9V403C125STM)
which has a fast readout (up to 200 full frames per second),
but is relatively noisy (80 electron rms/pixel). The cameras have
a trigger signal for camera synchronization and a firewire (IEEE 1394a) bus 
for data I/O, with 8 bits of output selectable within a 12-bit dynamic range.
The CMOS detector has 9.9 micron pixels in a 656 by 491 format. In operation we read out a 
312 by 312 section of the chip. With this region of interest, the maximum frame rate is
210 Hz, but in practice we operate between 100 and 128 Hz. After readout, the data are binned 
2x2 which results in sub-apertures separated by 10 pixels,
allowing for accurate centroiding and spot size measurements, and eliminating
sub-aperture boundary crossings. Because of the high read noise, the
wavefront sensors have an effective magnitude limit of 7.25 on the
6.5 meter Clay (Magellan 2) telescope. At maximum gain, variations in bias level within a 
frame are significant, but subtraction of a bias frame leaves the data flat to a few percent.
The peak efficiency of the detector is 32\% at 550 nm, and the system
has an effective wavelength of 620 nm. 

The Basler A602f camera follows an industry standard for firewire
cameras (version 1.30 of the \char`\"{}1394 - based Digital Camera
Specification\char`\"{} (DCAM) issued by the 1394 Trade Association).
This allowed us to write our own control software on a linux-based
PC using the open-source implementation of the DCAM specifications
(libdc1394).
\begin{sidewaystable}
\begin{center}
\begin{tabular}{|c||c|c|c|c|c|c|c|c|c|}
\hline 
Run&
Date&
JD&
\# Cameras&
Separations&
Clay&
DIMM&
MASS&
Wind Speed&
Temp\tabularnewline
\hline 
&
&
&
&
(arcminutes)&
(arcsec)&
(arcsec)&
(arcsec)&
(mph)&
(C)\tabularnewline
\hline
\hline 
1&
06/04/04&
2453161.821&
2&
7.04&
0.43&
N/A&
N/A&
13&
11.8\tabularnewline
\hline 
2&
09/23/04&
2453271.526&
2&
13.85&
0.96&
N/A&
N/A&
17&
13.3\tabularnewline
\hline 
3&
09/24/04&
2453272.552&
2&
13.85&
0.69&
0.70&
0.70&
13&
16.4\tabularnewline
\hline 
4&
09/26/04&
2453273.501&
2&
13.85&
0.68&
0.70&
0.47&
12&
14.6\tabularnewline
\hline 
5&
09/27/04&
2453275.517&
2&
13.85&
0.76&
0.77&
0.51&
5&
12.8\tabularnewline
\hline 
6&
09/27/04&
2453275.740&
2&
14.88&
0.82&
0.79&
0.52&
3&
14.1\tabularnewline
\hline 
7&
09/27/04&
2453275.799&
2&
15.54&
1.09&
0.97&
1.21&
14&
13.9\tabularnewline
\hline 
8&
09/28/04&
2453276.724&
2&
14.88&
1.04&
N/A&
N/A&
15&
14.3\tabularnewline
\hline 
9&
09/28/04&
2453276.795&
2&
15.54&
0.81&
0.88&
0.95&
12&
14.4\tabularnewline
\hline 
10&
09/28/04&
2453276.845&
2&
12.90&
1.02&
1.10&
1.05&
8&
14.2\tabularnewline
\hline 
11&
05/30/05&
2453520.830&
4&
4.30, 7.35, 9.19, 9.41, 11.26, 15.60&
0.69&
N/A&
N/A&
7&
15.7\tabularnewline
\hline 
12&
05/31/05&
2453521.833&
3&
7.35, 9.41, 15.60&
0.55&
0.45&
0.24&
9&
16.4\tabularnewline
\hline 
13&
09/19/05&
2453632.908&
2&
6.0&
0.72&
N/A&
N/A&
31&
13.2\tabularnewline
\hline 
14&
09/20/05&
2453633.602&
3&
6.0, 7.7, 12.8&
0.99&
1.00&
0.52&
30&
11.1\tabularnewline
\hline 
15&
09/21/05&
2453634.513&
3&
6.0, 7.7, 12.8&
0.89&
0.89&
0.33&
29&
11.3\tabularnewline
\hline
\end{tabular}
\end{center}
\caption[Description of the engineering runs for the GLAO experiment at Magellan.]{Engineering 
  runs for the GLAO experiment at Magellan.}
\label{Obs_Table}
\end{sidewaystable}

Each of the 4 cameras is attached to a separate computer, with a fifth computer to provide
overall synchronization. The raw data rate is substantial at 0.5 Gigabyte/min/camera, but a 
few minutes of data are adequate to characterize the ground-layer behavior at any given time, 
and inexpensive hard disk drives provide adequate storage capacity.  Exposure start times are 
controlled through the use of a common trigger voltage sent to all of the cameras.  Time 
synchronization of the frames recorded by separate computers can be a problem because of local
clock drift between computers and occasional dropped frames.  For this reason a pattern of 
randomly spaced 1 and 2 millisecond pauses is imposed on the basic 100 Hz trigger signal, 
which establishes a unique time sync pattern which can be recovered by looking at the difference 
in time stamps between frames.  The hardware setup allowed for continuous
recording of data and all the frames were saved to disk for later
processing.

\section{Observing Runs}

We conducted five observing runs over two years (2004 and 2005), mostly 
using engineering time and sharing the telescope most nights with other 
tests or science observations.  The
first run was used to field test the hardware and work on the software.
The other four runs produced 11 nights of wide-field, 100 Hz wavefront
sensing data. Typically the setup would take half an hour per camera
and then once all the cameras were aligned, the wavefront sensing
data would be recorded for approximately ten minutes. Setup involved
initializing the hardware and then the aligning the pupil on the lenslet 
array in each wavefront sensor.

Table \ref{Obs_Table} lists the parameters of each data set. The
number of cameras and sampled baselines between the cameras are listed
in columns 4 and 5. The seeing from the Clay guider, DIMM and MASS
instruments are listed in columns 6-8. These seeing measurements are
the averages in the ten minutes prior to the observation end time
(listed as a Julian Date in column 3). All of the seeing measurements
have been corrected for airmass and to a reference wavelength of $0.5\,\mu m$.
The wind speed and temperature are reported as the average conditions
for ten minutes surrounding the observation.

\section{Reductions/Results}


The imaging data was corrected for bias offsets, binned two by two
and smoothed with a double pass 3 x 3 box car filter, which eliminates 
sharp peaks and cosmic rays and minimizes centroid errors. Centroids, with 
sub-pixel accuracy, were obtained by a linear interpolation around the 
brightest pixel in each sub-aperture. The results from this fast and simple 
centroiding method differed from a more precise, moment centroiding analysis 
by less than 0.08 pixels rms. A reference image was built from roughly six 
thousand frames (60 seconds of data) and used to define the zero-deviation, 
reference positions for the Shack-Hartmann analysis.

A 36-mode Zernike decomposition (complete up to 6th radial+azimuthal order) 
was performed on each instantaneous image from the Shack-Hartmann imaging. The 
end product of the data reduction is a time-series of Zernike coefficients for 
each camera. Typically 12,000 frames (2 minutes) of data are analyzed.  Note 
that the tip/tilt terms include both atmospheric effects and tracking errors of 
the mount, particularly wind-shake.  During the observations guide corrections 
were sent to the mount using the standard guide cameras in the usual way, with 
a response time of about one second.

\begin{table}
\begin{center}
\begin{tabular}{|c|c|c|c|}
\hline
Zernike Mode&
RMS&
Zernike Mode&
RMS\tabularnewline
\hline 
&
$(arcsec/\mu m)$&
&
$(arcsec/\mu m)$\tabularnewline
\hline
\hline 
2,3 (Tip/Tilt)&
0.13&
20, 21&
1.04\tabularnewline
\hline 
4 (Focus)&
0.63&
22&
2.74\tabularnewline
\hline 
5,6 (Astig)&
0.45&
23, 24&
2.59\tabularnewline
\hline 
7,8 (Coma)&
0.91&
25, 26&
2.13\tabularnewline
\hline 
9,10 (Trefoil)&
0.64&
27, 28&
1.26\tabularnewline
\hline 
11 (Spherical)&
1.50&
29, 30&
3.41\tabularnewline
\hline 
12, 13&
1.35&
31, 32&
3.18\tabularnewline
\hline 
14, 15&
0.84&
33, 34&
2.54\tabularnewline
\hline 
16, 17&
1.95&
35, 36&
1.47\tabularnewline
\hline 
18, 19&
1.73&
&
\tabularnewline
\hline
\end{tabular}
\caption[RMS per Micron Wavefront Error]{Image Diameter RMS (arcsec) per Micron of Wavefront Error.}
\label{Zernike-Mode-Table}
\end{center}
\end{table}

Since GLAO reduces the wavefront error but typically does not result in 
diffraction-limited images, it is useful to be able to consider the effect
of the Zernike modes as quadrature contributions to the RMS image diameter 
in arcseconds.  Higher-order modes produce larger images for a given wavefront 
error because the slope errors occur over shorter scales.  Table \ref{Zernike-Mode-Table}
presents a list of conversion factors expressed in arcseconds per micron of wavefront
error. The table was derived using Zernike phase screens in a Zemax model of a perfect 6.5 
meter telescope.  Note that the RMS image diameter is determined from the image centroid,
which in the case of coma and higher-order coma may neglect some of the error which shows 
up as a shift in the image centroid.

Table 3 is a summary of the properties of the Zernike analysis of each run. 
Column 2 is the seeing as observed through the 59 cm sub-apertures. This number is
derived from the total wavefront error (col 3), the fractional wavefront errors 
observed in each Zernike order (col 4-10) and the conversion factors given in Table 2.

\begin{sidewaystable}
\begin{center}
\begin{tabular}{|c|c|c|c|c|c|c|c|c|c|}
\hline 
Run&
Seeing&
$rms(Z_{total}$)(nm)&
$\frac{<Z_{tip/tilt}^{2}>}{<Z_{total}^{2}>}$&
$\frac{<Z_{order=1}^{2}>}{<Z_{total}^{2}>}$&
$\frac{<Z_{order=2}^{2}>}{<Z_{total}^{2}>}$&
$\frac{<Z_{order=3}^{2}>}{<Z_{total}^{2}>}$&
$\frac{<Z_{order=4}^{2}>}{<Z_{total}^{2}>}$&
$\frac{<Z_{order=5}^{2}>}{<Z_{total}^{2}>}$&
$\frac{<Z_{order=6}^{2}>}{<Z_{total}^{2}>}$\tabularnewline
\hline
\hline 
1&
0.37&
1100&
0.519&
0.297&
0.066&
0.055&
0.026&
0.025&
0.012\tabularnewline
\hline 
2&
0.78&
1520&
0.780&
0.119&
0.039&
0.036&
0.016&
0.008&
0.002\tabularnewline
\hline 
3&
0.72&
1520&
0.810&
0.107&
0.032&
0.029&
0.013&
0.006&
0.002\tabularnewline
\hline 
4&
0.62&
1240&
0.791&
0.117&
0.036&
0.031&
0.015&
0.008&
0.002\tabularnewline
\hline 
5&
0.69&
1380&
0.782&
0.123&
0.038&
0.033&
0.015&
0.008&
0.002\tabularnewline
\hline 
6&
0.66&
1070&
0.706&
0.146&
0.055&
0.053&
0.022&
0.013&
0.004\tabularnewline
\hline 
7&
0.87&
1680&
0.770&
0.127&
0.040&
0.036&
0.016&
0.008&
0.002\tabularnewline
\hline 
8&
1.39&
2610&
0.782&
0.113&
0.039&
0.036&
0.016&
0.009&
0.004\tabularnewline
\hline 
9&
0.75&
1380&
0.752&
0.135&
0.044&
0.040&
0.018&
0.009&
0.003\tabularnewline
\hline 
10&
0.86&
1780&
0.800&
0.112&
0.035&
0.031&
0.014&
0.007&
0.002\tabularnewline
\hline 
11&
0.64&
1060&
0.706&
0.152&
0.053&
0.049&
0.023&
0.012&
0.004\tabularnewline
\hline 
12&
0.41&
860&
0.734&
0.139&
0.048&
0.044&
0.021&
0.011&
0.003\tabularnewline
\hline 
13&
0.72&
1850&
0.882&
0.060&
0.023&
0.021&
0.009&
0.005&
0.001\tabularnewline
\hline 
14&
1.17&
2330&
0.799&
0.105&
0.038&
0.034&
0.014&
0.008&
0.001\tabularnewline
\hline 
15&
0.61&
1130&
0.767&
0.122&
0.042&
0.039&
0.017&
0.009&
0.002\tabularnewline
\hline
\hline 
Average&
0.75&
1500&
0.756&
0.134&
0.043&
0.038&
0.017&
0.009&
0.003\tabularnewline
\hline
\end{tabular}
\caption[Modal Content]{Total Wavefront Aberrations and Modal Contribution to Total.}
\end{center}
\label{Modal_Table}
\end{sidewaystable}

\begin{figure}
   \begin{center}
   \begin{tabular}{c}
   \includegraphics[height=9cm]{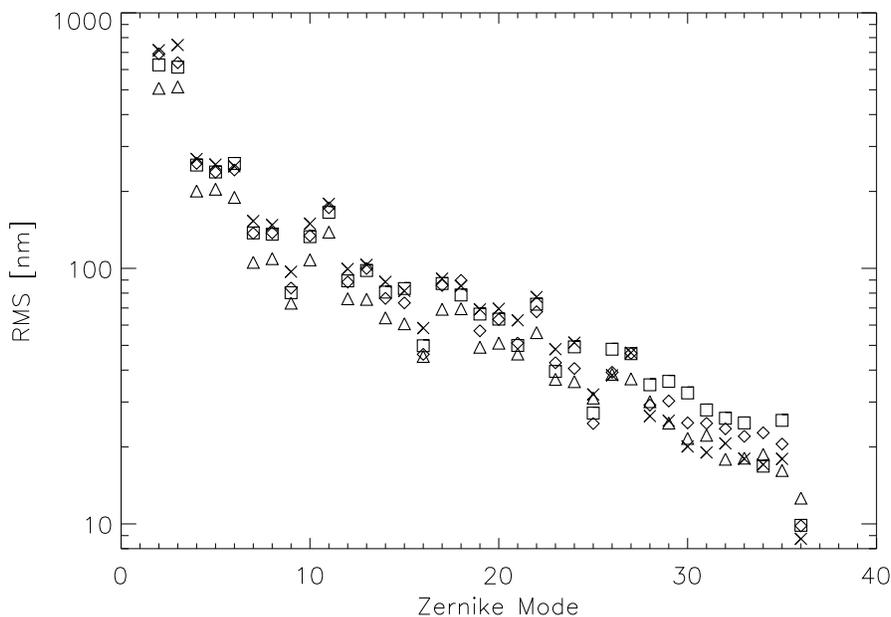}
   \end{tabular}
   \end{center}
   \caption[example] 
   { \label{rms_nm} 
     Example of the RMS of the Zernike modes in nm of wavefront error.  The 
     four different symbols represent the four cameras from Run 11.
   }
   \end{figure}

Figure \ref{rms_nm} shows an example of the RMS of the Zernike amplitudes for Run
11. The overall shape of the RMS as a function of mode is predicted
by atmospheric theory and is useful for deriving measurements of $r_{o}$,
the outer scale, and testing the validity of Kolmogorov turbulence (to be done
in a future study in combination with full presentation of seeing
monitor data). Note that there is some variation in the overall RMS from camera to camera. 
This is due partly to fainter stars having noisier measurements, but also partly
to small variations in the spacing of the optics in the wavefront sensors. The optical 
effect also includes focus errors caused by the curvature of the telescope focal plane, 
although we attempt to correct for this during the setup for each observation.  To move 
the measurements on a common scale, the values for each wavefront sensor are scaled to 
the average for all wavefront sensors in a data set.

\begin{figure}
   \begin{center}
   \begin{tabular}{c}
   \includegraphics[height=9cm]{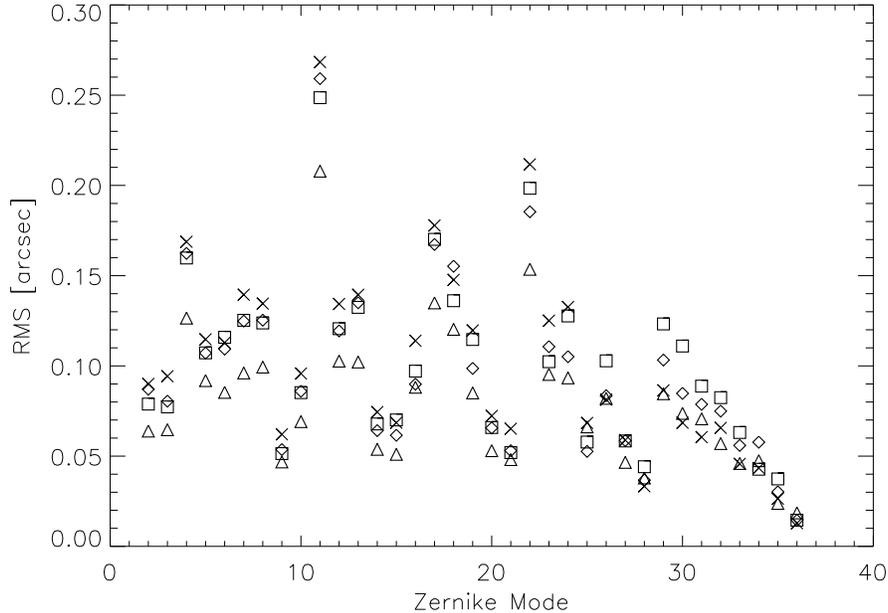}
   \end{tabular}
   \end{center}
   \caption[example] 
   { \label{rms_arcsec} 
     Example of the contribution to the total image diameter of the Zernike modes in arcsec.  The 
     four different symbols represent the four cameras from Run 11.
   }
   \end{figure}

The RMS as a function of Zernike mode is plotted in arcseconds in
Figure \ref{rms_arcsec}. Most of the modes contribute comparably to the total seeing,
although it is clear that the contribution is falling off after mode
30 and the radial terms (4, 11, 22) contribute more than the average
mode. Also note that although 75\% of the wavefront variance is contained
in tip/tilt (Table 3), the contribution to RMS image
size from these two modes is quite small. 

\begin{figure}
   \begin{center}
   \begin{tabular}{c}
   \includegraphics[height=9cm]{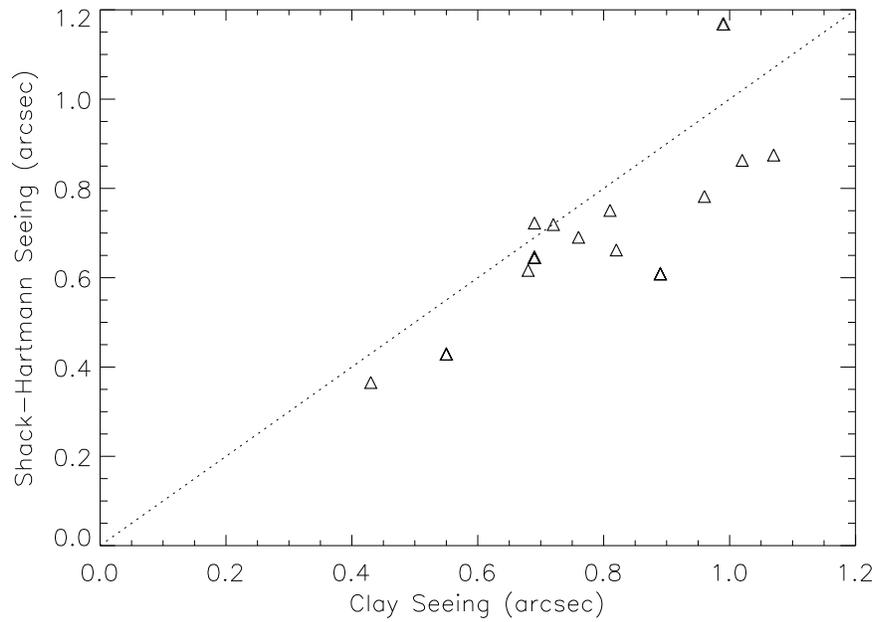}
   \end{tabular}
   \end{center}
   \caption[example] 
   { \label{seeing_weR_seeing} 
     The Clay seeing is plotted against the seeing detected in the
     59 cm sub-apertures of the GLAO experiment.  The dotted line
     represents the 1\:1 relation.  All seeing measurements have 
     been corrected to a reference wavelength of $0.5 \mu m$.  The 
     59 cm sub-apertures detect 86\% of the total atmospheric variance.
   }
   \end{figure}

Figure \ref{seeing_weR_seeing} shows the seeing from the Clay guide camera compared with
the seeing determined from the wavefront sensors. On average, the
59 cm sub-apertures of our wavefront sensors detect 86\% of the variance
of the total seeing. This implies that any fractional ground-layer contribution
that we measure applies only to the 86\% of the seeing that we detect. Presumably 
the {}``missing seeing'' is in higher order modes, which will
have small isoplanatic scale lengths and little impact in a GLAO system.

\section{GLAO Analysis}

For each pair of cameras, we decompose the wavefront errors in each Zernike 
mode into a correlated component, $Z_{G}$, which we attribute to the ground-layer
and an uncorrelated component, $Z_{U}$, which we attribute to the upper 
atmosphere. Any noise associated with the system is also attributed to 
$Z_{U}$, since the upper atmosphere is effectively noise for a system 
attempting to sense the ground-layer. Then for wavefront sensors 1 and 2, 
the wavefront errors are $Z_{1}=Z_{G}+Z_{U,1}$ and $Z_{2}=Z_{G}+Z_{U,2}$, where $Z_{U,\mathbf{\#}}$
is the upper atmosphere turbulence seen through each wavefront sensor
and $Z_{G}$ is common to both cameras by definition. Define
$P\equiv Z_{1}+Z_{2}$, and $M\equiv Z_{1}-Z_{2}$. The variance of $P$ is 
$<P^{2}>=4<Z_{G}^{2}>+2<Z_{U}^{2}>$ and the
variance of $M$ is $<M^{2}>=2<Z_{U}^{2}>$. The fraction of the variance arising 
from the ground-layer is 
$GL\equiv\frac{<P^{2}>-<M^{2}>}{<P^{2}>+<M^{2}>}=\frac{<Z_{G}^{2}>}{<Z_{G}^{2}>+<Z_{U}^{2}>}=\frac{Ground-Layer\, Variance}{Total\, Variance}$. 

In Table 4 we report the fraction of atmospheric wavefront
variance seen in the ground-layer as a function of Zernike radial
order. Where multiple baselines were observed within a run, each is
reported as a separate line. Tip and tilt are not distinct from mount
shake and may not reflect the ground-layer seeing. In the high wind
data, Runs 13-15, a clear elongation is seen of the spots, indicating
some mount shake. In the next section we put these eleven nights of ground-layer
into context and convert these fractional variance measures into image
size. As a consistency check, two observations from separate nights
were analyzed with this method. These data which are known to be completely
uncorrelated produced results of $GL<0.02$ for all orders. We note
that Run 8 is very close to zero and not consistent with the other
runs in any of the modes. We are in the process of reexamining this
dataset more closely.

\begin{figure}
   \begin{center}
   \begin{tabular}{c}
   \includegraphics[height=9cm,angle=90]{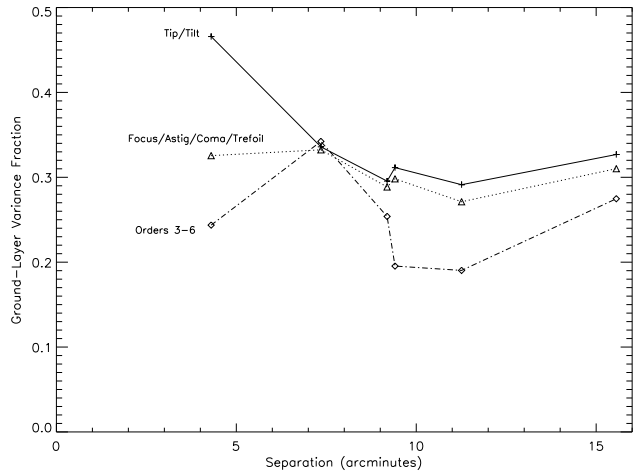}
   \end{tabular}
   \end{center}
   \caption[example] 
   { \label{gl_vs_sep} 
     Example of the field effect of the ground-layer.  The six 
     baselines are plotted for Run 11.  Modes are averaged together 
     to provide increased signal-to-noise.
   }
   \end{figure}

In Figure \ref{gl_vs_sep} the fraction of variance in the ground-layer, $GL$, 
is plotted against separation for Run 11. The expected behavior is that the ground-layer
contribution will fall off as field angle to the $5/3$ power. However the ground-layer 
fraction appears to fall off more slowly or even remain approximately constant as field 
angle increases. 

\begin{sidewaystable}
\begin{center}
\begin{tabular}{|c|c|c|c|c|c|c|c|c|c|}
\hline 
Run&
Separation&
$GL_{tip/tilt}$&
$GL_{order=1}$&
$GL_{order=2}$&
$GL_{order=3}$&
$GL_{order=4}$&
$GL_{order=5}$&
$GL_{order=6}$&
All Modes Weighted\tabularnewline
\hline
\hline 
1&
7.04&
0.33&
0.24&
0.13&
0.14&
0.13&
0.10&
0.07&
0.112\tabularnewline
\hline 
2&
13.85&
0.49&
0.22&
0.23&
0.20&
0.17&
0.15&
0.13&
0.208\tabularnewline
\hline 
3&
13.85&
0.60&
0.20&
0.09&
0.04&
0.05&
0.06&
0.06&
0.100\tabularnewline
\hline 
4&
13.85&
0.60&
0.36&
0.29&
0.23&
0.20&
0.17&
0.14&
0.257\tabularnewline
\hline 
5&
13.85&
0.46&
0.25&
0.16&
0.15&
0.13&
0.12&
0.11&
0.172\tabularnewline
\hline 
6&
14.88&
0.27&
0.23&
0.27&
0.35&
0.33&
0.29&
0.29&
0.308\tabularnewline
\hline 
7&
15.54&
0.31&
0.15&
0.13&
0.10&
0.09&
0.09&
0.10&
0.116\tabularnewline
\hline 
8&
14.88&
0.02&
0.03&
0.04&
0.05&
0.03&
0.04&
0.02&
0.039\tabularnewline
\hline 
9&
15.54&
0.21&
0.19&
0.19&
0.17&
0.19&
0.16&
0.15&
0.182\tabularnewline
\hline 
10&
12.90&
0.18&
0.08&
0.06&
0.07&
0.03&
0.05&
0.04&
0.068\tabularnewline
\hline 
11&
4.30&
0.46&
0.36&
0.29&
0.29&
0.26&
0.18&
0.13&
0.272\tabularnewline
\hline 
&
7.35&
0.34&
0.34&
0.32&
0.38&
0.35&
0.30&
0.21&
0.336\tabularnewline
\hline 
&
9.19&
0.31&
0.30&
0.30&
0.25&
0.19&
0.15&
0.09&
0.227\tabularnewline
\hline 
&
9.41&
0.30&
0.30&
0.28&
0.30&
0.26&
0.20&
0.13&
0.279\tabularnewline
\hline 
&
11.26&
0.29&
0.31&
0.24&
0.25&
0.19&
0.13&
0.08&
0.219\tabularnewline
\hline 
&
15.56&
0.32&
0.31&
0.31&
0.33&
0.28&
0.22&
0.14&
0.291\tabularnewline
\hline 
12&
7.35&
0.44&
0.42&
0.34&
0.34&
0.34&
0.28&
0.24&
0.336\tabularnewline
\hline 
&
9.41&
0.46&
0.33&
0.24&
0.21&
0.19&
0.13&
0.10&
0.217\tabularnewline
\hline 
&
15.86&
0.35&
0.28&
0.23&
0.25&
0.22&
0.16&
0.12&
0.226\tabularnewline
\hline 
13&
6.0&
0.75&
0.36&
0.16&
0.17&
0.10&
0.07&
0.02&
0.198\tabularnewline
\hline 
14&
6.00&
0.72&
0.66&
0.44&
0.39&
0.38&
0.26&
0.20&
0.449\tabularnewline
\hline 
&
7.70&
0.62&
0.45&
0.24&
0.19&
0.12&
0.07&
0.06&
0.231\tabularnewline
\hline 
&
12.90&
0.74&
0.39&
0.17&
0.09&
0.01&
0.02&
$<$0.01&
0.157\tabularnewline
\hline 
15&
6.00&
0.79&
0.70&
0.60&
0.54&
0.49&
0.35&
0.24&
0.535\tabularnewline
\hline 
&
7.70&
0.83&
0.72&
0.57&
0.54&
0.44&
0.32&
0.18&
0.519\tabularnewline
\hline 
&
12.90&
0.69&
0.64&
0.54&
0.50&
0.46&
0.33&
0.21&
0.497\tabularnewline
\hline
\end{tabular}
\caption[Fraction of Ground-Layer]{Fraction of Variance Detected in Ground-Layer}
\end{center}
\label{GL_Table}
\end{sidewaystable}

\section{Supporting Data}

In order to put the GLAO experiment data into context, we have examined
a year's worth of site monitoring data. Meteorological data are being
collected at the Magellan site using weather stations manufactured
by Davis Instruments Corp. A Weather Monitor II station has been in
continuous operation at site since late-2000. In 2005 this
station was replaced with two newer models (Vantage Pro \& Vantage
Pro2) which utilize the same basic sensor technology to measure temperature,
humidity, atmospheric pressure, and wind speed and direction. The
stations generate measurements for temperature, pressure, humidity,
wind speed, and wind direction every 1-2 seconds and are stored in
a database as 1 minute averages. 

The seeing is being monitored at the Magellan telescopes through the
use of a differential image motion monitor (DIMM). The DIMM \cite{DIMM} 
functions by using Kolmogorov turbulence theory to relate
the FWHM of a long exposure in a large telescope to the variance in
the difference in the positions of two images of the same star created 
by placing two weak prisms at the aperture of a small telescope. The Magellan 
DIMM is based on the CTIO RoboDIMM, but has several improvements. Following a technique developed
by the TMT project, software was written to use a SBIG ST7 CCD in
a drift scan readout mode which allows for many more image motion
measurements to be made per minute and thus improved statistics. Image
quality has also been improved by using two thinner prisms as opposed
to one thicker prism and an open aperture. 

   \begin{figure}
   \begin{center}
   \begin{tabular}{c}
   \includegraphics[height=9cm]{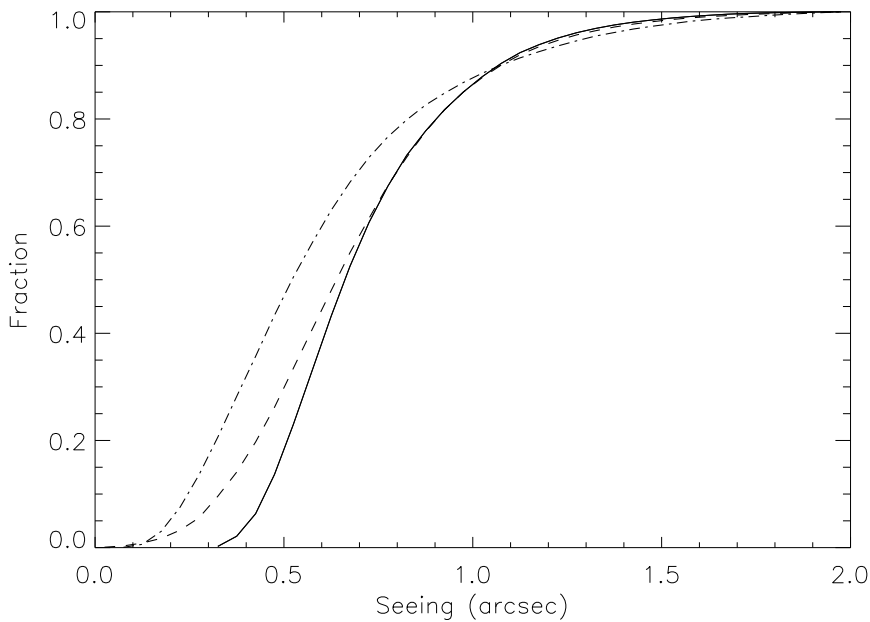}
   \end{tabular}
   \end{center}
   \caption[example] 
   { \label{cmd_hist} 
     Cumulative seeing histogram for Clay (solid), DIMM (dash),
     MASS (dot-dash) for one-year's worth of data.  See Table
     \ref{MDC_Quartiles} for quartile seeing values.
   }
   \end{figure}

Turbulence in the atmosphere above 500 m is being monitored by a multi-aperture
scintillation sensor (MASS, \cite{MASS}). The spatial scale of the
scintillation variation depends on the distance to the turbulence which gives rise 
to the wavefront phase disturbance.
The turbulence distribution in a small number of discrete layers
can be determined by fitting a model to the differences between the
scintillation indices within four concentric apertures. Our MASS has
an accompanying DIMM built into it. This instrument, known as a MASS-DIMM,
was fabricated and provided by CTIO and put into operation in a tower
at the Magellan telescopes site. The MASS-DIMM instrument operates
at the back-end of a 0.25 meter Meade telescope in its own, ventilated
10-meter tower. The MASS is only sensitive to turbulence above 500
m. In addition to the turbulence profile, the MASS also measures free
atmospheric seeing (essentially the integral of the turbulence profile),
the adaptive optics time constant and the isoplanatic angle. The difference
between the DIMM seeing and the free atmosphere MASS seeing is a measure
of the seeing contributed by the ground layer. 

In addition to the site monitoring equipment, the seeing from both
Magellan Telescopes is recorded automatically. The Magellan guide
cameras have low quality optics with significant aberrations and so
a correction (0.35'' in quadrature) must be removed from the observed
FWHM values. The guide cameras use a RG610 filter and that combined
with the midband coated E2V CCD47-20 produces an effective wavelength
of $765\, nm$.

We examined 205 nights from March 2004 to April 2005 that had both
seeing measurements from all the instruments and meteorological data.
(Note that the clear fraction of nights is higher, but the MASS-DIMM
instrument was being commissioned in the first half of 2004.) For
each seeing measurement, a mean temperature, pressure, wind speed
and wind direction was defined as the mean meteorological condition
for 15 minutes surrounding the seeing measurement. In the end, we
assembled 64,000 MASS and 64,000 DIMM seeing measurements and 128,000
Clay guide camera seeing measurements. All of the data were corrected
to an airmass of 1, and to a reference wavelength of $0.5\mu m$.
The MASS data were reprocessed with the most recent version of the
turbina software.

The first consistency check is to compare the Magellan seeing measurements
to the DIMM seeing. In Figure \ref{cmd_hist} the cumulative seeing histograms are
plotted for the Clay guide camera (solid), DIMM (dash), and MASS (dot-dash)
and in Table \ref{MDC_Quartiles} select values are presented numerically.
The agreement between the DIMM and Clay is good, but not perfect.
The low end difference between the two is likely caused by the simple
correction applied to compensate for the guide camera optics. This
correction is a first order correction and could likely be improved
with further analysis. The effects of an outer scale will suppress
the seeing in the larger Clay Telescope aperture. However, since the
DIMM is a differential measure (and a much smaller structure), the
DIMM seeing values will be unaffected by mount and wind shake.  It appears that
to first order, these two effects cancel each other out and we obtain good agreement
between the Magellan guiders and the DIMM.  
\begin{table}
\begin{center}
\begin{tabular}{|c|c|c|c|}
\hline 
\label{MDC_Quartiles}Percentile&
MASS&
DIMM&
Clay\tabularnewline
\hline
\hline 
10\%&
0.25&
0.33&
0.45\tabularnewline
\hline 
25\%&
0.35&
0.47&
0.53\tabularnewline
\hline 
50\%&
0.52&
0.64&
0.66\tabularnewline
\hline 
25\%&
0.75&
0.85&
0.84\tabularnewline
\hline 
10\%&
1.07&
1.07&
1.07\tabularnewline
\hline
\end{tabular}
\caption[Seeing Quartiles]{Seeing Percentiles for MASS, DIMM, and Clay}
\end{center}
\end{table}

The MASS data relative to the DIMM data indicates that in the median
conditions, ground-layer is responsible for about 38\% of the atmospheric
variance. This is a smaller percentage of total seeing than is observed
at other sites \cite{Pachon_DIMM,MASS_2,MASS_3,2002MNRAS.337..103W}.
At Las Campanas, a ground-layer correction would work in the sense
of making a good night better as there is little difference seen between
the MASS and DIMM at higher seeing values.

   \begin{figure}
   \begin{center}
   \begin{tabular}{c}
   \includegraphics[height=12cm,angle=90]{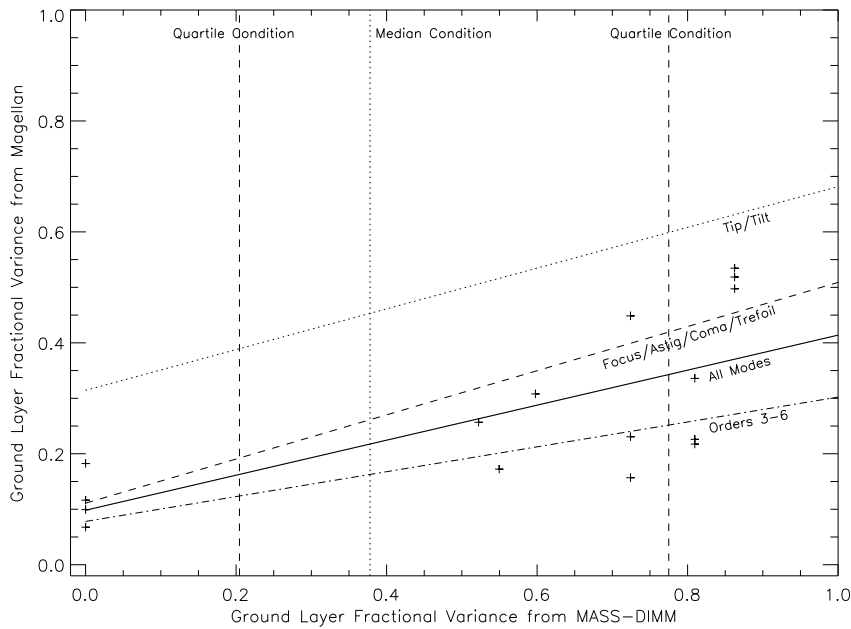}
   \end{tabular}
   \end{center}
   \caption[example] 
   { \label{gl_md} 
     MASS-DIMM measured ground-layer vs ground-layer from this experiment.
     The mean relation between ground-layer variance as measured by the MASS-DIMM instrument
     and the Magellan Telescope
     is plotted as a solid line and labeled All Modes.
     The data for All Modes are plotted as crosses to provide a sense
     of the errors.  
     Trend-lines for tip/tilt (dot), Focus, Astigmatism, Coma, Trifoil (dashed), and higher
     orders (dot-dash) are drawn as a fit to the data from Table \ref{GL_Table}.
     The vertical lines indicate the quartile conditions as determined
     from a year's worth of seeing data.
   }
   \end{figure}

In Figure \ref{gl_md} the ground-layer as measured by MASS-DIMM, $(1-\frac{MASS^{2}}{DIMM^{2}})$, 
is plotted against the trends lines for ground-layer fractional variance
as measured by our experiment. We set any MASS-DIMM data that results in
a negative value to zero and attribute this to noise and small errors in
the absolute calibration between the two instruments. There is a correlation
in between the MASS-DIMM and the GLAO experiment, although the relationship
is far from one-to-one. The crosses are the data that were used to determine
the trend-line for orders 3-6 and provide a sense of the uncertainties.
It is interesting to note that even when the MASS-DIMM indicates that
there is no ground-layer, there is a significant tip/tilt correlation,
indicating that approximately 30\% of the variance in tip/tilt is
due to mount shake. 

There is also a small offset in the higher order modes.  Since a GL 
measurement between two separate nights produced a result of less than 0.02, the 
positive values of about 0.1 in the higher-order modes, even when the MASS-DIMM comparison
fails to detect the ground layer, are likely to be significant.

   \begin{figure}
   \begin{center}
   \begin{tabular}{c}
   \includegraphics[height=12cm,angle=90]{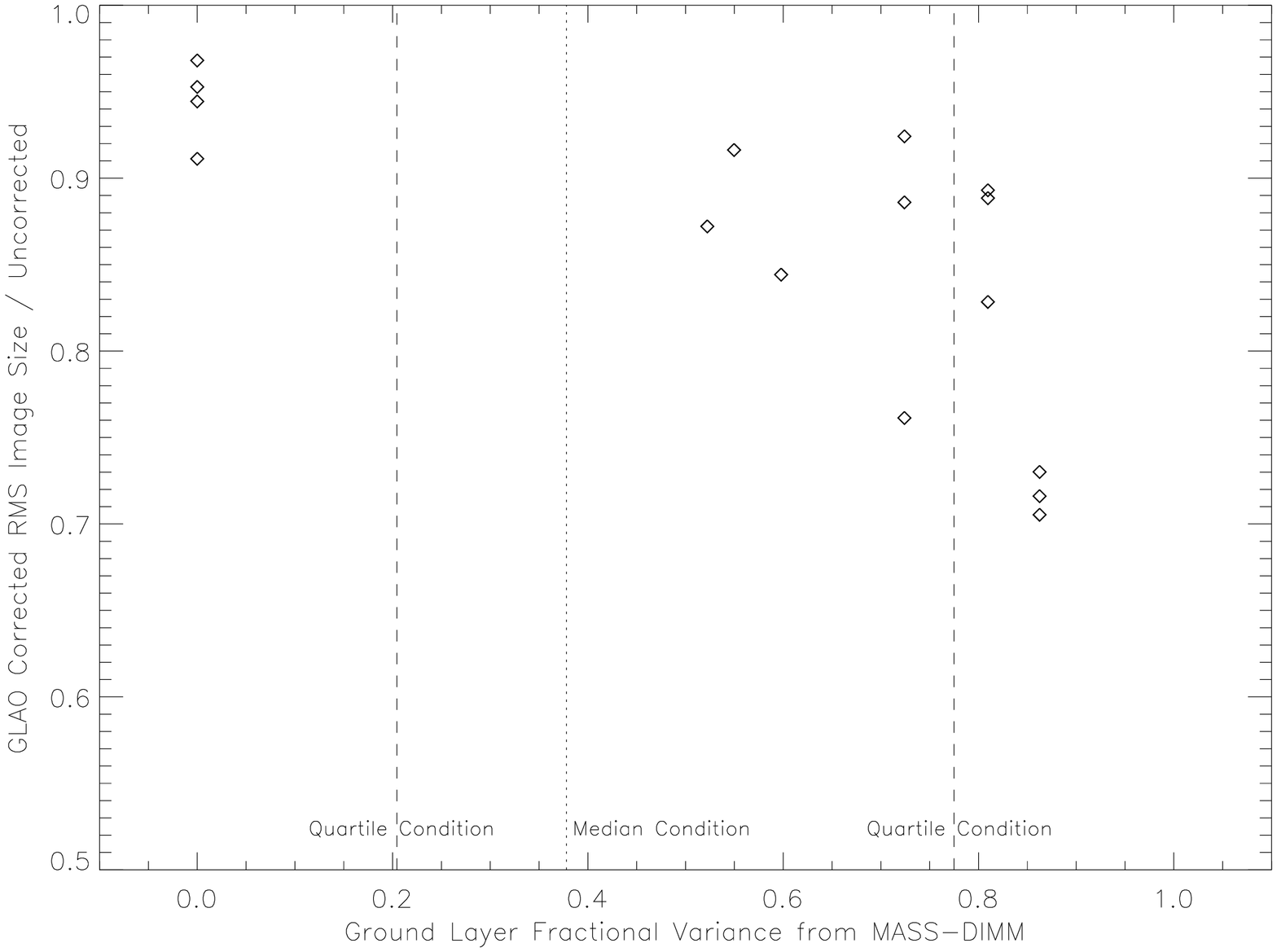}
   \end{tabular}
   \end{center}
   \caption[example] 
   { \label{rms_md} 
     Improvement in RMS images diameter shown as a function of MASS-DIMM reported ground-layer.
     Vertical lines indicate the quartile and median conditions.
     Note that not all the runs have MASS-DIMM data, and each baseline for a given run is plotted, 
     providing multiple measurements for some nights.  This calculation includes a correction
     to account for the fact that 86\% of the total seeing variance is observed in the 59 cm sub-apertures.
   }
   \end{figure}

Figure \ref{rms_md} shows the ground-layer as measured by MASS-DIMM and the RMS
image improvement. These numbers were derived by reducing each mode
in the RSS of the total variance by the amount of observed ground-layer
reduction (Table 4). These scaled variances are then
converted to arcseconds and compared to the uncorrected RSS variance
from the same data. These calculations include the scaling
factor that accounts for the fact that the 59 cm sub-apertures only
see 86\% of the seeing. 

The largest gains are obtained when the MASS-DIMM reports the strongest ground-layer.
Note that multiple points will show up for a
given observation if more than one baseline was observed. It is clear
that strong ground-layer nights allow for improvements in image quality,
although the amount of improvement is significantly less than predicted.
Neither of the two previous figures are changed significantly by the
removal of the largest ($>$10 arcmin) baseline data. The strongest ground-layer
night also happened to be the night of the strongest winds (and technically
over the operating limit!). However, no correlation exists between
image correction and ground wind speed.

\section{Conclusion}

We have taken one of the first sets of moderate speed (100 Hz), wide-field
(arcminutes), and moderate resolution (59 cm sub-aperture) wavefront
sensor data, in order to asses the gains of a GLAO system. We have
taken this data on eleven nights which span a range of meteorological,
seeing, and reported ground-layer strength conditions. On the best
nights, we find a RMS image size reduction of 30\% over a 7 arcminute
separation. On a typical night we find a 10\% correction to a visible-band
image.  

\acknowledgments 
This work has been supported by the Seaver Foundation. We are very grateful
to I. Thompson for sharing telescope time with this project. We would
also like to thank the staff of Las Campanas.

\bibliography{glao}  
\bibliographystyle{spiebib}   

\end{document}